# Achieving Efficient and Secure Data Acquisition for Cloud-supported Internet of Things in Smart Grid

Zhitao Guan, Jing Li, Longfei Wu, Yue Zhang, Jun Wu, Xiaojiang Du

*Abstract*—**Cloud-supported Internet of Things (Cloud-IoT) has been broadly deployed in smart grid systems. The IoT front-ends are responsible for data acquisition and status supervision, while the substantial amount of data is stored and managed in the cloud server. Achieving data security and system efficiency in the data acquisition and transmission process are of great significance and challenging, because the power grid-related data is sensitive and in huge amount. In this paper, we present an efficient and secure data acquisition scheme based on CP-ABE (Ciphertext Policy Attribute Based Encryption). Data acquired from the terminals will be partitioned into blocks and encrypted with its corresponding access sub-tree in sequence, thereby the data encryption and data transmission can be processed in parallel. Furthermore, we protect the information about the access tree with threshold secret sharing method, which can preserve the data privacy and integrity from users with the unauthorized sets of attributes. The formal analysis demonstrates that the proposed scheme can fulfill the security requirements of the Cloud-supported IoT in smart grid. The numerical analysis and experimental results indicate that our scheme can effectively reduce the time cost compared with other popular approaches.**

*Index Terms*—**Cloud-supported IoT, smart grid, CP-ABE, data acquisition, parallel.**

## I. INTRODUCTION

With the support of modern information technologies like the Internet of Things (IoT) and cloud computing, smart grid has emerged as the next-generation power supply network, in which the electricity is generated according to the real-time demands of electric equipment or household appliances [1,2]. To make the smart grid more intelligent, a great number of IoT terminals are deployed to gather the status of the power grid timely for the control center. Some sample applications are shown in Fig. 1, such as the power transmission line monitoring, power generation monitoring, substation state monitoring, smart metering, electric energy data acquisition, smart home. For instance, in power transmission line monitoring scenario, using preplaced sensors, the status parameters of the transmission line and power towers can be gathered in real time, so that any fault can be diagnosed and located in a timely manner.

In smart grid, the different kinds of applications mentioned above all generate an enormous amount of data, which needs to be stored and managed efficiently. Cloud-IoT is proposed to address this issue [2,3]. As shown in Fig. 1, with the support of cloud computing, mass data from different IoT terminals can be collected and processed by local front-end servers, then transferred and stored in the cloud servers. The data in cloud can be accessed by various types of data users. The power grid staff can continually monitor the status of power grid. Researchers and government agencies can analyze the data for research or policymaking.

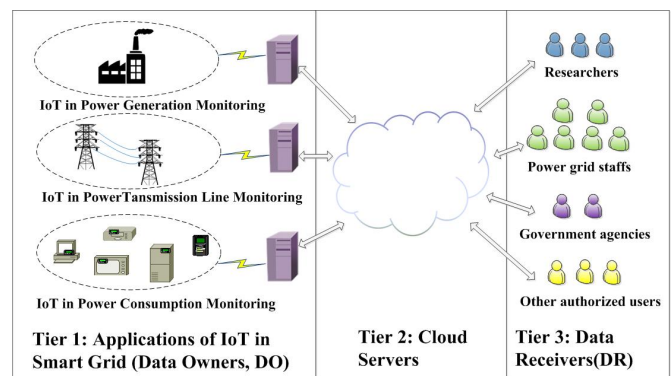

Fig. 1 Illustration of cloud-supported IoT in smart grid

Actually, some works on Wireless Sensor Networks can be used for reference, such as [4-9]. However, there are still several problems and challenges in smart grid data acquisition. First, the efficiency of data acquisition should be considered due to the large amount of data to be encrypted/decrypted and transferred. It's critical to ensure an acceptable the data acquisition time. Second, the protection of data security and privacy must be kept in mind. To deal with these two problems

*Jing Li is the corresponding author.

Zhitao Guan, Jing Li and Yue Zhang are with School of Control and Computer Engineering, North China Electric Power University, Beijing, 102206, China (e-mail: guan@ncepu.edu.cn, li_jing@ncepu.edu.cn, minemail@yeah.net).

Longfei Wu and Xiaojiang Du are with the Department of Computer and Information Sciences, Temple University, Philadelphia, USA (e-mail: longfei.wu@temple.edu, dxj@ieee.org).

Jun Wu is with the College of Information Security Engineering, Shanghai Jiao Tong University, Shanghai, China (e-mail: junwuhn@sjtu.edu.cn).

simultaneously, in this paper, we present an efficient and secure data acquisition scheme based on CP-ABE. The main contributions of our work can be summarized as the following:

⬦ We propose a parallel data processing method. Data acquired from the terminals will be partitioned into blocks and encrypted with its corresponding access sub-tree in sequence, thereby the data encryption and data transmission can be processed in parallel. The data decryption process is similar to the process of data encryption.

⬦ We introduce the dual secret sharing scheme to protect the access tree information. Only when all of the shares are combined can the secret be recovered. Each of the data blocks holds a share. While the last one share is protected with the other secret sharing scheme. If the user's attributes satisfy the threshold function of root node, then the last share will be retrieved. In addition, some users with the unauthorized attributes sets will be filtered out. We realize the privacy-preserving, the data integrity check and the attributes check simultaneously.

⬦ We give the security analysis and performance evaluation, which prove that the security of our scheme is no weaker than that of the traditional scheme, and that our scheme can reduce the system response time and users' waiting time notably.

The rest of this paper is organized as follows. Section 2 introduces the related work. The preliminaries are given in section 3. In section 4, the system model and security model are described. In section 5, we present the details of the proposed scheme. Its security analysis and performance evaluation are conducted in section 6 and section 7, respectively. Section 8 concludes the paper.

## II. RELATED WORK

Recently, various techniques have been proposed to address the problems of data security and fine-grained access control. In [10], Sahai and Waters proposed the Attribute-Based Encryption (ABE) to realize fine-grained access control on encrypted data. In ABE, the encryption policy is associated with a set of attributes, and the data owner can be offline after data is encrypted. Vipul Goyal et al developed a new cryptosystem for fine-grained sharing of encrypted data in [11] based on Sahai's work, called Key-Policy Attribute-Based Encryption (KP-ABE). In their scheme, the ciphertext's encryption policy is associated with a set of attributes, but the attributes that organized into a tree structure (named access tree) are specified by data receivers. In [12], Bethencourt et al proposed the Ciphertext Policy Attribute Based Encryption (CP-ABE). In their work, the data owner constructs the access tree using visitors' identity information. The user can decrypt the ciphertext only if the attributes in his private key match the access tree.

Owing to the particular advantages of ABE, they are often applied to protect the outsourced data. Yu et al tried to achieve secure, scalable, and fine-grained access control in a cloud environment [13]. Their proposed scheme was based on KP-ABE, and combined with another two techniques, proxy re-encryption and lazy re-encryption. It was proved that the proposed scheme can meet the security requirements quite well in cloud. Similarly, [14-16] are other applications of KP-ABE in cloud. Han et al [16] defined a weak anonymity of ABE scheme and proposed a general transformation from ABE to Attribute Based Encryption with Keyword Search.

In [17], the researchers proposed the first KP-ABE scheme enabling truly expressive access structures with constant ciphertext size. In [18], Li et al proposed an expressive decentralizing KP-ABE scheme with constant ciphertext size, which allows the access policy to be expressed as any non-monotone access structure. Similarly, in [19], a fully secure scheme in the standard model is proposed, which is with constant-size ciphertexts and fast decryption simultaneously. There are some other similar researches, such as [20]. The comparison of the above schemes is showed in Table I.

TABLE I.    THE COMPARISON OF SCHEMES

|  | Constant ciphertext size | Access Structure | KP-ABE/ CP-ABE | Security |
|---|---|---|---|---|
| [17] | √ | Non-monotonic | KP-ABE | CCA-secure |
| [18] | √ | Non-monotonic | KP-ABE | Semantic secure |
| [19] | √ | Monotonic | KP-ABE | Fully secure in Standard oracle |
| [20] | √ | Monotonic | CP-ABE | CCA-secure |

To improve system efficiency and protect the user privacy, some researchers study on the multiple authorities. Fu et al [21] firstly presented the Attribute-based Encryption (ABE), secure deletion and secret-sharing schemes and construed a secure multi-authority access control scheme. The scheme proposed in [22] could reduce the reliance on the central authority and protect users' privacy. The proposed scheme allows each authority to work independently without any collaboration to initialize the system and issue secret keys to users. In [23], Li et al presented a low complexity multi-authority attribute based encryption scheme for mobile cloud computing, which uses masked shared-decryption-keys to ensure the security of decryption outsourcing. They adopted multi-authorities for authorization, to enhance the security assurance.

How to ensure the integrity and correctness of data are challenging issues. In [24], Yadav and Dave presented an access model based on CP-ABE which could provide a remote integrity check by way of augmenting secure data storage operations. Zhou adopted a similar method [25]. In addition to the access tree division, Zhou also proposed an efficient data management model to balance the communication and storage overhead and reduce the cost of data management operations. Lai et al [26] gave the formal model of ABE with verifiable outsourced decryption and propose a concrete scheme. Lin et al [27] proposed a generic construction of verifiable outsourced ABE (VO-ABE), based on an attribute-based key encapsulation mechanism (AB-KEM). It can be considered in both key-policy (KP) and ciphertext-policy (CP) settings. Mao et al [28] proposed generic constructions of CPA-secure and RCCA-secure ABE systems with verifiable outsourced decryption from CPA-secure ABE with outsourced decryption, respectively.

In [29], M Green et al proposed a proxy re-encryption in Identity-Based Encryption scheme, in which the ciphertexts can be transformed from one identity to another. In [30], R Canetti et al proposed a definition of security against chosen ciphertext attacks for PRE schemes. An IBE PRE scheme was proposed in [31] without random oracles. In [32], the first unidirectional proxy re-encryption scheme was proposed with chosen-ciphertext security in the standard model. In [33], SSM Chow et al's scheme gains high efficiency and CCA security using the "token-controlled encryption" technique. H Wang et al realized an identity-based proxy re-encryption scheme that can achieve IND-CCA2 secure [34]. In contrast with requiring users to be online all the time, a time-based proxy re-encryption (Time-PRE) scheme was presented in [35] to revoke a user's access right automatically after a predetermined period of time. In [36], the proxy re-encryption was proved to be useful as a method of adding access control to the SFS read-only file system. This scheme also achieves a stronger notion of security.

## III. Preliminaries

### A. Bilinear Maps

Let $G_0$ and $G_1$ be two multiplicative cyclic groups of prime order $p$ and $g$ be the generator of $G_0$. The bilinear map $e$ is, $e : G_0 \times G_0 \to G_1$, for all $a, b \in \mathbb{Z}_p$:

- Bilinearity: $\forall u, v \in G_1, e(u^a, v^b) = e(u, v)^{ab}$.
- Non-degeneracy: $e(g, g) \neq 1$.
- Symmetric: $e(g^a, g^b) = e(g, g)^{ab} = e(g^b, g^a)$.

### B. Discrete Logarithm (DL) Problem:

Let $G$ be a multiplicative cyclic group of prime order $p$ and $g$ be its generator. Given a tuple $< g, g^x >$, where $g \in_R G$ and $x \in \mathbb{Z}_p$ are chosen as input uniformly at random, the DL problem is to recover $x$.

**Definition 1** The DL assumption holds in $G$ is that no probabilistic polynomial-time (PPT) algorithm $\mathcal{A}$ can solve the DL problem with negligible advantage. We define the advantage of $\mathcal{A}$ as follows:

$$\Pr[\mathcal{A} < g, g^x >= x]$$

The probability is over the generator $g$, randomly chosen $x$ and the random bits consumed by $\mathcal{A}$.

### C. Structure in Ciphertext-policy Attribute Based Encryption (CP-ABE)

**Definition 2** Let $P = \{P_1, P_2, ..., P_n\}$ be a set of participants, let $U = 2^{\{P_1, P_2, ..., P_n\}}$ be the universal set. If $\exists AS \subseteq U \setminus \{\varnothing\}$, then $AS$ can be viewed as an access structure. If $A \in AS, \forall B \in U, A \subseteq B$, and $B \in AS$, $AS$ is considered as a monotonic access structure. Then the sets in $AS$ are defined as authorized sets, while the other sets are regarded as unauthorized sets.

To achieve fine-grained access control, we utilize the Ciphertext Policy Attribute-Based Encryption scheme. For ease of partition, we adopt the Bethencourt's scheme in [12], in which the access structure is illustrated by an access tree.

Let $\mathcal{T}$ be an access tree, and the root node is denoted by $\mathcal{R}$. All the leaves represent the attributes, while the interior nodes represent the threshold gates, described by its children and a threshold value, such as AND ($n$ of $n$), OR (1of $n$), and $n$ of $m$ ($m > n$).

At the beginning of the encryption, we randomly choose a secret $s$ and conduct a polynomial for each node from top to bottom, while the decryption order is reverse.

Additionally, some functions are necessary to be introduced. We use ***parent***$(x)$ to get the parent of the node $x$. The function ***att***$(x)$ is used only if $x$ is a leaf node and returns the attribute associated with the leaf node $x$ in the tree. The function ***index***$(x)$ returns the number associated with the node $x$, where the index values are uniquely assigned to nodes in the tree.

To retrieve the secret, we define the Lagrange coefficient $\Delta_{i,S}$ as follows:

For $i \in \mathbb{Z}_p$, and for $\forall x \in S$,

$$\Delta_{i,S(x)} = \prod_{j \in S, j \neq i} \frac{x - j}{i - j}.$$

### D. (t, n) Threshold Secret Sharing

Secret sharing scheme is used for sharing a secret among a group of parties, each of whom only obtain a piece of the secret (namely a share of the secret). No single party can infer any information about the secret with its own share. The only way to reconstruct the secret is to combine a certain number of shares.

The most basic secret sharing scheme is $(t, n)$ threshold scheme, which was first proposed by Shamir [37]. In his scheme, a secret divided into $n$ parts can be recovered only if at least t parts are collected. This idea has already been used to implement the tree-access structure.

## IV. DEFINITIONS

### A. Definition of System Model

Our system consists of four entities, namely Data Owner, Cloud Server, Attribute Authority, and Data Requester/Receiver. Both Data Owners (denoted as $DO$) and Data Requester/Receivers (denoted as $DR$) are users.

**Data Owners ($DO$)** $DO$ decide the access policy and encrypt the data with CP-ABE. The encrypted data will be uploaded to the Cloud Servers. $DO$ are assumed to be honest in the system.

**Data Requester/Receivers ($DR$)** $DR$ send the decryption request to Cloud and obtain the ciphertexts over the internet. Only when their attributes satisfy the access policies of the ciphertext, can they get access to the plaintexts. Data requester/receivers may collude to access the data that is otherwise not accessible individually.

**Cloud Servers ($CS$)** $CS$ are responsible for storing a massive volume of data. They cannot be trusted by $DO$. Hence, it is necessary for $DO$ to define the access policy to ensure the data confidentiality. CS are assumed not to collude with $DR$.

**Attribute Authority (AA)** AA is responsible for registering users, evaluating their attributes and generating their secret key $SK$ accordingly. It runs the *Setup* algorithm, and issues public key $PK$ and master key $MK$ to each $DO$. It is considered as fully trusted.

### B. Definition of Our system

**Definition 3** Our scheme consists of the following algorithms: *Setup*, *Key_Gen*, *Encryption*, and *Decryption*.

**Setup** $(1^\lambda) \rightarrow PK, MK$. The setup algorithm takes a security parameter $\lambda$ as input and outputs the public key and master key.

**Key_Gen** $(PK, MK, S) \rightarrow SK$. The key generation algorithm takes the public key $PK$, the master key $MK$ and a set $S$ of attributes that belongs to users as inputs. It outputs a corresponding secret key $SK$.

**Encryption** $(PK, M, AS) \rightarrow CTB$. The encryption algorithm consists of two subroutines: data partition subroutine **Data_Partition** and data block encryption subroutine **DB_Encryption**.

- **Data_Partition** $(M, AS) \rightarrow DB_1...DB_n$. This subroutine takes the message M and access structure $AS$ as inputs. It outputs a number of encrypted data blocks $DBs$. The number is decided by the threshold function.

- **DB_Encryption** $(DB_i, AS_i, PK, MK) \rightarrow CTB_i$. The data block encryption subroutine takes the data blocks $DBs$, the corresponding sub-access-structure $AS_i$, public key $PK$ and master key $MK$ as inputs. It outputs the ciphertext blocks $CTBs$.

**Decryption** $(DB_i, AS_i, PK, SK) \rightarrow DB_i$. The data decryption algorithm consists of six subroutines: check of user's attributes subroutine **ATT_Check**, check of the ciphertext blocks integrity subroutine **CTB_Integrity**, the leaf nodes decryption subroutine **Decrypt_LeafNode**, the interior nodes decryption subroutine **Decrypt_InteriorNode**, two ciphertext block decryption subroutines **CTB_ABE_Dec** and **CTB_SYM_Dec**.

- **ATT_Check** $(CTB_n, S) \rightarrow R_{n+1}$ or $\perp$. This subroutine takes as inputs the last ciphertext block $CTB_n$ and user's attributes set $S$. If the check is passed, it returns $R_{n+1}$. Otherwise, it outputs $\perp$.

- **CTB_Integrity** $(CTB.id, R_{n+1}) \rightarrow \mathcal{T}$ or $\perp$. This subroutine takes as inputs all the id of ciphertext blocks and the output of **ATT_Check**. If the check is passed, it outputs the access tree information $\mathcal{T}$. Otherwise, it outputs $\perp$.

- **Decrypt_LeafNode** $(CTB_i, SK, z, \mathcal{T}) \rightarrow F_z$ or $\perp$. This subroutine takes as inputs ciphertext blocks, user's secret key, leaf nodes information and the access tree $\mathcal{T}$. It outputs $F_z$ or $\perp$.

- **Decrypt_InteriorNode** $(F_z, S_i, \mathcal{T}) \rightarrow F_i$ or $\perp$. This subroutine takes as inputs the outputs of **Decrypt_LeafNode**, the leaf nodes set $S_i$ and the access tree $\mathcal{T}$. If $S_i$ satisfies the current sub-access-policy, it outputs $F_i$. Otherwise, it outputs $\perp$.

- **CTB_ABE_Dec** $(CTB_i, SK, F_i) \rightarrow DB_i$ or $\perp$. This subroutine takes as inputs $CTB_i$, user's $SK$, and the output of **Decrypt_InteriorNode**. It outputs $DB_i$ or $\perp$.

- **CTB_SYM_Dec** $(CTB_i, E_j, SK) \rightarrow DB_i$ or $\perp$. This subroutine takes as inputs the $CTB_i$, user's $SK$ and a parameter $E_j$. It outputs $DB_i$ or $\perp$.

### C. Definition of Security Model

Here, we introduce the universal security model of our system, which is defined similar to Bethencourt's scheme in [12]. In this security model, there is an adversary $\mathcal{A}$ and a challenger $\mathcal{C}$. The adversary is allowed to challenge on an encryption to the access structure $AS^*$ and query for any secret keys $SKs$ as long as they cannot be applied directly to decrypt the ciphertext. The challenger is responsible for the ciphertext generation under the $AS*$ and the secret key generation. Now the security game is described as follows:

**Setup:** The challenger $\mathcal{C}$ runs this algorithm. It gives the public parameters PK to the adversary $\mathcal{A}$ and keeps MK to itself.

**Phase 1:** $\mathcal{A}$ issues queries for repeated private keys corresponding to the sets of attributes $S_1,...S_{q_1}$. ($q$ and $q_1$ are integers randomly chosen by $\mathcal{A}$, with $1 < q_1 < q$ ). If any of the sets $S_1,...S_{q_1}$ satisfies the access structure $AS*$, then aborts. Else, $\mathcal{C}$ generates the corresponding secret keys to the sets for $\mathcal{A}$. Then he submits a set of attributes and a ciphertext $CT$, and obtains the corresponding $M$ from $\mathcal{C}$.

**Challenge:** $\mathcal{A}$ submits two equal length messages $M_0$ and $M_1$ to $\mathcal{C}$. The challenger $\mathcal{C}$ randomly flips a coin $b$, and encrypts $M_b$ under the challenge access structure $AS*$. Then the generated ciphertext $CT*$ will be given to $\mathcal{A}$.

**Phase 2:** Repeat **Phase 1**, and the sets are turned from $S_1,...S_{q_1}$ to $S_{q_1+1},...S_q$.

**Guess:** The adversary $\mathcal{A}$ outputs its guess $b' \in \{0,1\}$ for $b$ and wins the game if $b' = b$ .

The advantage of an adversary $\mathcal{A}$ in this game is defined as

$$Adv(\mathcal{A}) = \left| \Pr[b' = b] - \frac{1}{2} \right|,$$

where the probability is taken over the random bits used by the challenger and the adversary.

**Definition 4** A CP-ABE scheme is CCA-secure if polynomial time adversaries have at most a negligible advantage in the above game.

**Definition 5** A CP-ABE scheme is CPA-secure if the adversaries cannot make decryption queries in Phase 1.

## V. DESCRIPTION OF OUR SYSTEM

### A. Overview

We partition the file and the access tree into blocks, which allows each data block to be encrypted and decrypted independently, in parallel to the transmission and computation of the different data blocks. This reduces the response time of severs and shortens the $DR$'s waiting time.

As soon as a block is encrypted, it is ready to be transmitted, which allows the next block to be encrypted at the same time. As shown in Fig.2, we use $T_1$ to denote the total time of encryption and transmission of our scheme and $T_2$ denote the time of CP-ABE. The diagram below shows the advantage of our scheme.

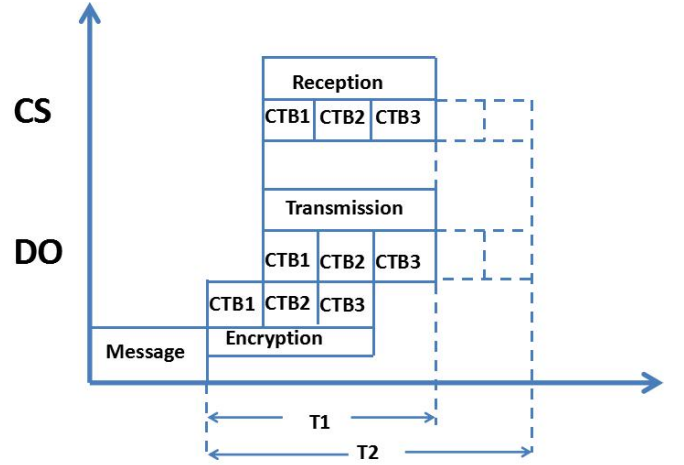

Fig. 2. The encryption-transmission time comparison between our scheme and CP-ABE

As shown in Fig.3, the similar situation occurs in decryption phase as well. DR decrypts one block, while others are being transmitted over the internet. We use $T_1$ to denote the total time of transmission and decryption of our scheme and $T_2$ denote the time of CP-ABE. The diagram shows the advantage of our scheme.

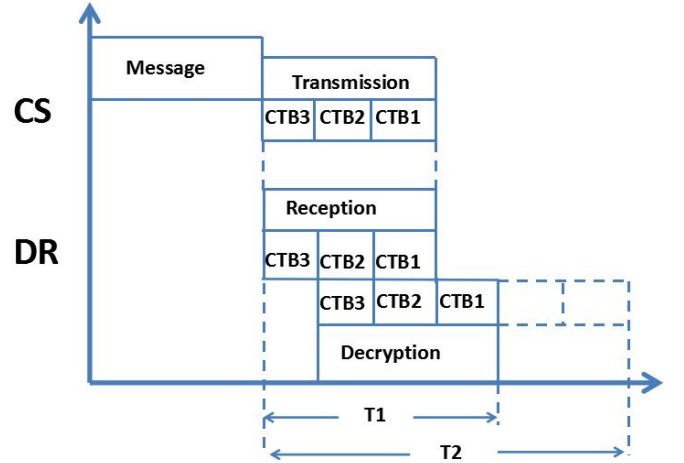

Fig. 3. The transmission-decryption time comparison between our scheme and CP-ABE

### B. Algorithms

Here we will give the details of the algorithms in our system.

#### 1) Setup

This setup algorithm will choose a bilinear group $G_0$ of prime order $p$ with generator $g$. Then, it will choose three random exponents：$\alpha, \beta, q \in \mathbb{Z}_p$. The public key and the master key are published as:

$$PK = \left\{ G_0, g, h = g^\beta, e(g,g)^\alpha \right\} \tag{1}$$

$$MK = \left\{ \beta, g^\alpha, q \right\} \tag{2}$$

#### 2) Key_Gen

This operation is implemented by a specific trusted server independently. It is responsible for registering legal $DR$s, evaluating the attributes sets and generating the secret keys $SK$s accordingly. The algorithm is as follows:

**Key_Gen** ($PK$, $MK$, $S$) $\rightarrow SK$

This algorithm takes as inputs the set of attributes S, public parameters $PK$ and master keys $MK$. A corresponding secret key is the output. This algorithm first chooses $r_j \in \mathbb{Z}_P$, $\forall j \in S$ at random for each attribute. The Hash function $H_{att} : \{0,1\}^* \rightarrow G_0$ is introduced to map any attribute described as a binary string to a random group element. The $SK$ is computed as:

$$SK = \begin{pmatrix} D = g^{\frac{\alpha+r}{\beta}}, \forall j \in S, \hat{D} = g^{rq} \\ D_j = g^r H_{att}(j)^{rj}, D_j{}' = g^{rj} \end{pmatrix} \qquad (3)$$

*3) Encryption*

This process consists of two main algorithms, Data_Partition and DB_Encryption. The former one is used for partitioning the original data into several data blocks, the latter one is responsible for encrypting these data blocks one by one, which conducts the encryption and transmission in parallel.

**Data_Partition** ($M$, $n$) $\rightarrow DB_1 \dots DB_n$

DO first constructs an access tree for the given access policy and gets its interior node number $n$ (including the root node). This algorithm takes as inputs M and $n$, partitioning as follows:

$$M \rightarrow M_1, M_2, M_3, \dots, M_{n-1}, M_n \qquad (4)$$

$$\begin{aligned} &DB_1 = M_1, \\ &DB_2 = M_1 \oplus M_2, \\ &DB_3 = M_2 \oplus M_3, \\ &\dots, \\ &DB_n = M_{n-1} \oplus M_n \end{aligned} \qquad (5)$$

**DB_Encryption** ($DB_i$, $AS_i$, $PK$, $MK$) $\rightarrow CTB_i$

If $i=1$, namely that the first block is being encrypted under the first sub-tree. The sub-tree contains only two levels: the parent node and its child nodes. The algorithm performs like following:

It first selects $s \in \mathbb{Z}_p$ and $s_1 \in \mathbb{Z}_p$ at random and computes:

$$E_1 = g^{\frac{s_1}{q}} \qquad (6)$$

$$C_1 = (DB_1 \parallel g^{\frac{s_1}{q}}) e(g,g)^{\alpha s} \qquad (7)$$

$$C_1 = h^s \qquad (8)$$

According to the threshold of the root node, it will construct a polynomial $q_R(x)$, where $q_R(0) = s$, which is similar to Bethencourt's scheme in [12]. If there is any leaf node connecting to the root node, let $Y_1$ be the set of leaf nodes, it is computed as follows:

$$\forall y \in Y_1, \hat{C}_{1,y} = g^{q_y(0)}, \hat{C}_{1,y}{}' = H_{att}(att(y))^{q_y(0)} \qquad (9)$$

Where $q_y(0) = q_R(index(y))$.

In addition, if there is any child node $y'$ being the interior node of the tree, $q_{y'}(0) = q_R(index(y'))$. It will be the input for the next sub-tree construction.

Then, a random number $R_1$ generated by a pseudo-random generator will be considered as the $id$ of $CTB_1$.

$$CTB_1.id = R_1 \qquad (10)$$

The complete form of the first ciphertext block ($CTB_1$) is as follows:

$$CTB_1 = \left( C_1, C_1, \forall y \in Y_1, \hat{C}_{1,y}, \hat{C}_{1,y}{}' \right) \qquad (11)$$

If $i > 1$, namely that the first data block has been encrypted and transmitted over the internet and the other data blocks are ready for encrypting and transmitting, the encryption process is similar to that of the first data block. It selects $s_i \in \mathbb{Z}_p$ at random and computes:

$$C_i = (DB_i \parallel g^{\frac{s_i}{q}}) e(g,g)^{\alpha q_i(0)} \qquad (12)$$

$$C_1 = h^{q_i(0)} \qquad (13)$$

If there is any leaf node belonging to this sub-tree, let $Y_i$ be the set of leaf nodes, it is computed as follows:

$$\forall y \in Y_i, \hat{C}_{i,y} = g^{q_y(0)}, \hat{C}_{i,y}{}' = H_{att}(att(y))^{q_y(0)} \qquad (14)$$

The difference is that $q_i(0)$ denotes the parent node of the current sub-tree, where

$$j = parent(i), q_i(0) = q_j(index(i)),$$
$$i = parent(y), q_y(0) = q_i(index(y)).$$

To guarantee that each of the $CTB$ can be decrypted, the algorithm computes a difference value for this data block:

$$\Delta C_i = g^{\frac{s_j - q_i(0)}{q}} = g^{\Delta s_i} \qquad (15)$$

Similarly, set a random number $R_i$ to denote the $id$ of $CTB_i$.

$$CTB_i.id = R_i \qquad (16)$$

The complete form of the ciphertext block $CTB_i$ is as follows:

$$CTB_i = \left( C_i, C_i, \Delta C_i, \forall y \in Y_i, \hat{C}_{i,y}, \hat{C}_{i,y}{}' \right) \qquad (17)$$

If $i=n$, all the relevant computations about $CTB_n.id, C_n, C_n, \Delta C_n, \forall y \in Y_n, \hat{C}_{n,y}, \hat{C}_{n,y}'$ is similar to the previous operation. In addition, the complete information of the access tree is implicit and protected with $(n+1, n+1)$ secret sharing scheme by randomly generating $R_i$ and setting

$$R_{n+1} = \mathcal{T} \oplus R_1 \oplus R_2 \oplus R_3 ... \oplus R_n. \tag{18}$$

Assume that the threshold function of the root node is $k$ out of $t$, then we adopt $(k, t)$ secret sharing scheme to protect $R_{n+1}$. On one hand, the access policy information is implicit. On the other hand, users with the unauthorized attribute sets will be eliminated.

As we know, the root node $\mathcal{R}$ holds $t$ branches, and each branch holds several attributes, only when more than $k$ branches are satisfied can the secret $s$ be recovered. Thus, all attributes can be divided into $t$ disjoint sets: $Set_1, Set_2, ..., Set_t$.

The algorithm first constructs a polynomial $Q(x)$ whose degree is $k$-1 and $Q(0)= R_{n+1}$. Then it will select $t$ pairs of $(x_i, y_i)_{1 \leq i \leq t}$ and assign them to $t$ attribute sets. We introduce a one-dimensional array $A[x]$ and compute as follows:

If $att_i \in Set_1$, set

$$A[H_a(att_i)] = H_a(att_i) \oplus (x_1 \| y_1)$$

If $att_j \in Set_2$, set

$$A[H_a(att_j)] = H_a(att_j) \oplus (x_2 \| y_2)$$

...

If $att_k \in Set_t$, set

$$A[H_a(att_k)] = H_a(att_k) \oplus (x_t \| y_t)$$

Finally, it outputs:

$$CTB_n = (C_n, C_n, \Delta C_n, \forall y \in Y_n, \hat{C}_{n,y}, \hat{C}_{n,y}', A) \tag{19}$$

### 4) Decryption

Once the decryption request is sent, $DR$'s attributes need to be checked beforehand.

**ATT_Check** ($CTB_n, S$) $\rightarrow R_{n+1}$ or $\perp$

This algorithm takes as inputs the last ciphertext block and user's attribute set. It can filter out some users with unauthorized sets, which improves the decryption efficiency. It computes as follows:

$$\forall att_i \in S, A[H_a(att_i)] \oplus H_a(att_i) = x_j \| y_j \tag{20}$$

Finally, it may obtain $m$ pairs of $(x_j, y_j)_{1 \leq j \leq m}$.

With the polynomial interpolation, if $m \geq k$, the algorithm can recover the secret and output $R_{n+1}$; else, it outputs $\perp$.

**CTB_Integrity** ($CTB.id, R_{n+1}$) $\rightarrow \mathcal{T}$ or $\perp$

Next, the integrity of ciphertext will be checked by DR before entering the transmission-decryption mode. This algorithm takes as inputs the $id$ of all the ciphertext blocks and the output of **ATT_Check**. It computes as follows:

If all ciphertext blocks are available,

$$\begin{aligned} CTB_1.id \oplus CTB_2.id ... &\oplus CTB_n.id \oplus R_{n+1} \\ &= R_1 \oplus R_2 \oplus R_3 ... \oplus R_n \oplus R_{n+1} \\ &= \mathcal{T} \end{aligned} \tag{21}$$

Else, this algorithm outputs $\perp$.

After passing the checks for the attributes and integrity, it will enter the transmission-decryption mode. This algorithm takes as inputs the ciphertext block CTB and SK.

As each sub-tree only has two levels, there will be no recursion algorithm within the calculation. To complete the computation, we define another two algorithms Decrypt_LeafNode and Decrypt_InteriorNode.

**Decrypt_LeafNode** ($CTB_i$, $SK, z, \mathcal{T}$)$\rightarrow F_z$ or $\perp$

If the node $z$ is a leaf node, then we let $j = \textbf{att}(z)$ and perform the computation follows:

If $j \in S$, then

$$\begin{aligned} F_z &= \frac{e(D_j, \hat{C}_{i,z})}{e(D_j', \hat{C}_{i,z}')} \\ &= \frac{e(g^r H_{att}(j)^{rj}, g^{q_z(0)})}{e(g^{rj}, H_{att}(att(z))^{q_z(0)})} \\ &= e(g, g)^{r q_z(0)} \end{aligned} \tag{22}$$

Else, it returns $\perp$.

All of the leaf nodes contained in this $CTB_i$ will be calculated, and the outputs will be stored.

**Decrypt_InteriorNode** ($F_z, S_i, \mathcal{T}$)$\rightarrow F_i$ or $\perp$

If the node $i$ is the root node of a sub-tree, for all its child nodes $z$, the algorithm **Decrypt_LeafNode** ($CTB_i, SK, z, \mathcal{T}$) will compute them and store the outputs as $F_z$. Let $S_i$ be the set of child nodes $z$, if $\forall z \in S_i, F_z = \perp$ and the function returns $\perp$; else, we can compute the $F_i$ as follows:

$$\begin{aligned} F_i &= \prod_{z \in S_i} F_z^{\Delta_{z,S_i'}(0)}, where \begin{cases} n = index(z) \\ S_i' = \{index(z) : z \in S_i\} \end{cases} \\ &= \prod_{z \in S_i} (e(g, g)^{r \cdot q_z(0)})^{\Delta_{z,S_i'}(0)} \\ &= \prod_{z \in S_i} (e(g, g)^{r \cdot q_{parent(z)}(index(z))})^{\Delta_{z,S_i'}(0)} \\ &= \prod_{z \in S_i} (e(g, g)^{r \cdot q_i(n)})^{\Delta_{z,S_i'}(0)} \\ &= e(g, g)^{r \cdot q_i(0)} \end{aligned} \tag{23}$$

To guarantee that all of the *CTBs* can be decrypted even if some of the sub-tree cannot be satisfied，we propose two decryption methods:

**CTB_ABE_Dec** ($CTB_i$, $SK$, $F_i$) →$DB_i$ or ⊥

This subroutine takes as input the result of **Decrypt_InteriorNode**. If the result is valid, as long as one of them is retrieved (assume that the parent node $i$ is retrieved), this algorithm computes as follows:

$$\frac{C_i F_i}{e(C_i{}', D)} = \frac{(DB_i \parallel g^{\frac{s_i}{q}}) e(g,g)^{\alpha q_i(0)} e(g,g)^{r q_i(0)}}{e(h^{q_i(0)}, g^{\frac{\alpha+r}{\beta}})} \tag{24}$$

$$= (M_{i-1} \oplus M_i) \parallel g^{\frac{s_i}{q}}$$

**CTB_SYM_Dec** ($CTB_i$, $E_j$, $PK$, $MK$) →$DB_i$ or ⊥

Once there is one *CTB* being decrypted by satisfying its relevant sub-tree, the *CTBs* that are related to their child nodes can be decrypted legally.

Set $j = parent(i)$,

$$E = \frac{E_j}{\Delta C_i} = \frac{g^{\frac{s_j}{q}}}{g^{\Delta s_i}} = g^{\frac{s_j}{q} \cdot \frac{s_j - q_i(0)}{q}} = g^{\frac{q_i(0)}{q}}$$

$$\frac{C_i e(E, \hat{D})}{e(C_i, D)} = \frac{(DB_i \parallel g^{\frac{s_i}{q}}) e(g,g)^{\alpha q_i(0)} e(E, g^{rq})}{e(h^{q_i(0)}, g^{\frac{\alpha+r}{\beta}})}$$

$$= \frac{(DB_i \parallel g^{\frac{s_i}{q}}) e(g,g)^{\alpha q_i(0)} e(g,g)^{r q_i(0)}}{e(h^{q_i(0)}, g^{\frac{\alpha+r}{\beta}})} \tag{25}$$

$$= (M_{i-1} \oplus M_i) \parallel g^{\frac{s_i}{q}}$$

The message M can be retrieved as follows:

$$M_1 = M_1,$$
$$M_2 = DB_2 \oplus M_1,$$
$$M_3 = DB_3 \oplus M_2, \tag{26}$$
$$...$$
$$M_n = DB_n \oplus M_{n-1}$$

$$M = M_1 \parallel M_2 \parallel M_3 ... \parallel M_n \tag{27}$$

Finally, it outputs *M*.

## VI. SECURITY ANALYSIS

We analyze the security properties of our system considering the security model defined in Section 4.

### A. System Security

We parallel the transmission and computation by partitioning the data and access tree into chunks, which reduces the response time of severs and the *DR*'s waiting time. The ways to encrypt and decrypt will affect the security of the system.

**Theorem 1:** The security of our system is no weaker than that of [12].

**Proof:** We prove this theorem by the following game. Suppose that an adversary $\mathcal{A}$ can attack our scheme with non-negligible advantage.

- **Setup:** The challenger $\mathcal{C}$ runs this algorithm. It gives the public parameters *PK* to the adversary $\mathcal{A}$ and keeps *MK* to itself.

- **Phase 1:** $\mathcal{A}$ issues queries for the private keys corresponding to the sets of attributes $S_1, ... S_{q_1}$. In addition, he also submits an access structure *AS\**. If any of the sets satisfies the access structure *AS\**, then aborts. $\mathcal{C}$ generates the corresponding secret keys *SKs* to the sets for $\mathcal{A}$.

- **Challenge:** $\mathcal{A}$ submits two equal length messages $M_0$ and $M_1$ to $\mathcal{C}$. The challenger $\mathcal{C}$ randomly flips a coin $b$, and encrypts $M_b$ under the challenge access structure *AS\**. Then the generated ciphertext blocks *CTBs* will be given to $\mathcal{A}$.

- **Phase 2:** $\mathcal{A}$ issues queries for the private keys as in Phase 1, and the sets are turned from $S_1, ... S_{q_1}$ to $S_{q_1+1}, ... S_q$.

- **Guess:** The adversary $\mathcal{A}$ outputs its guess $b' \in \{0,1\}$ for $b$ and wins the game if $b' = b$.

Obviously, the game is formulated corresponding to the one in [12]. Thus, if $\mathcal{A}$ can attack our scheme with non-negligible advantage, he can attack the CP-ABE [12] as well.

### B. Partition

A complete tree will be partitioned into several sub-trees, and each data block (*DB*) will be encrypted with the CP-ABE. The difference is that each sub-tree only contains two levels: one root node and its child nodes. In case some ciphertext blocks are not decrypted, we propose two ways for decryption: one is the same as CP-ABE, the other one is to compute the root node value of current sub-tree from its parent block. However, it may lead to the consequence that even if the user's set doesn't satisfy the access policy, the user can still perform decryption and obtain some *DBs*.

**Theorem 2:** Partition doesn't impact the security of our scheme.

**Proof:** Compared with a complete tree, each level contains partial access policy. There is no other information inserted into these partitioned ones.

As for *DBs*, all of them are preprocessed for the first one. Without the first one, even all the rests are acquired, none of them can be retrieved. The only way to access the first $DB$ is to satisfy its access policy, which is the same as that of [12]. Thus, the partition of the message and the tree doesn't impact the security of the proposed scheme.

### C. Privacy-Preserving and Integrity

**Theorem 3:** Our scheme is secure against the adversaries with polynomial time in the length of the access tree information.

**Proof:** To protect the access tree information, we encrypt it with ($n+1$, $n+1$) threshold secret sharing scheme, that is, the secret will be shared among $n+1$ parties. Before encryption, each ciphertext block obtains a randomly generated string as its $id$. These strings are the n shares of the secret. The $n+1$-th share is implicit, encrypted with another ($k$, $t$) threshold secret sharing scheme.

Only with $n$ valid ciphertext blocks and the recovered n+1-th share, can the secret be retrieved. However, the strings of attributes are assigned to legal users by the trusted AA. Adversaries who have no knowledge about the attribute strings cannot launch the brute force attack to guess the attribute strings within polynomial time. Thus, they cannot detect the underlying information about the secret without attributes.

## IV. PERFORMANCE EVALUATION

### A. Numerical analysis

In this paper, we partition a large file into several data blocks, which can be encrypted, decrypted, and transmitted concurrently. Therefore, we are able to perform the encryption/decryption and the transmission of different blocks in parallel. Now, we give the efficiency analysis according to the process shown in Fig. 2 and Fig. 3. The time cost in each step is shown in Table II.

TABLE II.    THE PARAMETERS IN PERFORMANCE EVALUATION

| Symbols | Description |
| --- | --- |
| $n$ | Number of blocks M is partitioned into. |
| $ET_M$ | Encryption time of the whole file |
| $ET_i$ | Encryption time of data block i. |
| $TT_M$ | Transmission Encryption time of the whole file |
| $TT_i$ | Transmission time of data block i. |
| $DT_M$ | Decryption time of the whole file. |
| $DT_i$ | Decryption time of data block i. |

### 1) Encryption-Transmission

The total time of traditional scheme [12] is,

$$ET_M + TT_M \qquad (28)$$

If $TT_i > ET_i$, the total time of our scheme is,

$$ET_1 + \sum TT_i \approx ET_1 + TT_M \qquad (29)$$

Else, the total time is,

$$TT_n + \sum ET_i \approx ET_M + TT_n \qquad (30)$$

We set $\Delta T$ to denote the difference:

$$\Delta T \approx ET_M + TT_M - ET_1 - TT_M \approx ET_M - ET_1$$
$$\Delta T \approx ET_M + TT_M - ET_M - TT_n \approx TT_M - TT_n \qquad (31)$$

For all various data we suppose that there is no delay in the process of block encryption/decryption or transmission. Therefore, there is no time gap between two consecutive blocks in data block encryption, decryption, and transmission. Thus, in either case, $\Delta T > 0$.

### 2) Transmission-Decryption

The total time of traditional scheme [12] is,

$$TT_M + DT_M \qquad (32)$$

If $TT_i > ET_i$, the total time of our scheme is,

$$\sum TT_i + DT_n \approx TT_M + DT_n \qquad (33)$$

Else, the total time is,

$$TT_1 + \sum DT_i \approx DT_M + TT_1 \qquad (34)$$

We set $\Delta T$ denote the difference:

$$\Delta T \approx DT_M + TT_M - DT_n - TT_M \approx DT_M - DT_n$$
$$\Delta T \approx DT_M + TT_M - DT_M - TT_1 \approx TT_M - TT_1 \qquad (35)$$

Ideally, $DT_M > DT_n$, $TT_M > TT_1$.

Thus, in either case, $\Delta T > 0$.

### B. Experimental results

To evaluate the performance of our system, we implemented a testing environment using the cpabe-toolkit [38].

While most of the computation cost in encryption is due to the access tree, generating polynomials only takes a small portion of the time. We partition the data into blocks so that the encryption and transmission of data blocks can be executed simultaneously. In the comparative test, the different sizes of the text are first encrypted with the same access tree that contains ten levels and a hundred leaf nodes. As different decryption keys can affect the decryption time overhead, we use the same key to conduct the experiment Compared with [12], Fig. 4 (a) and Fig. 4 (b) show the difference of time cost between reference [12] and our scheme during encryption/decryption and transmission.

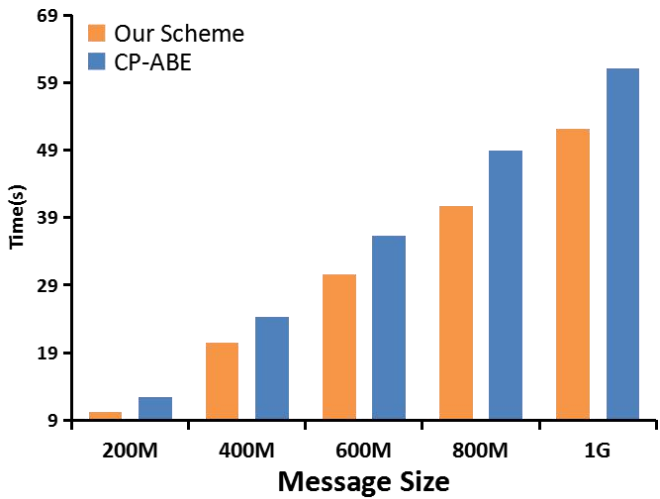

Fig.4 (a) Comparison of the encryption and transmission time between CP-ABE and our scheme, when the message size rises.

Encryption and transmission can be concurrently performed in data blocks. Compared to [12], the total time of encryption and transmission has been reduced.

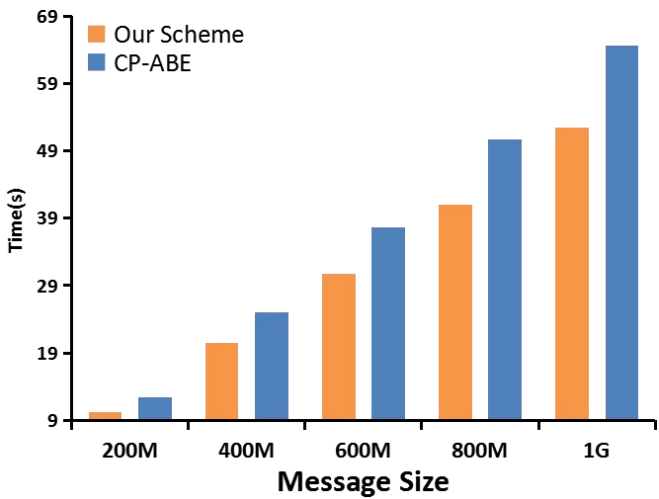

Fig. 4 (b) Comparison of the transmission and decryption time between CPABE and our system, when the number of blocks is different.

Similar to the encryption operation, the transmission and decryption are, also concurrently performed in data blocks. Compared to [12], the total time of decryption and transmission has been reduced.

Next, let the access trees share the same number of interior nodes (i.e., the number of blocks is fixed) but have different numbers of leaf nodes. Fig. 4 (c) shows the difference of time cost between reference [12] and our scheme during encryption. The total time of [12] rises as the number of leaf nodes grows. Although it increases as well in our scheme, the rate is very small and the time is significantly reduced, shown in Fig. 4 (c).

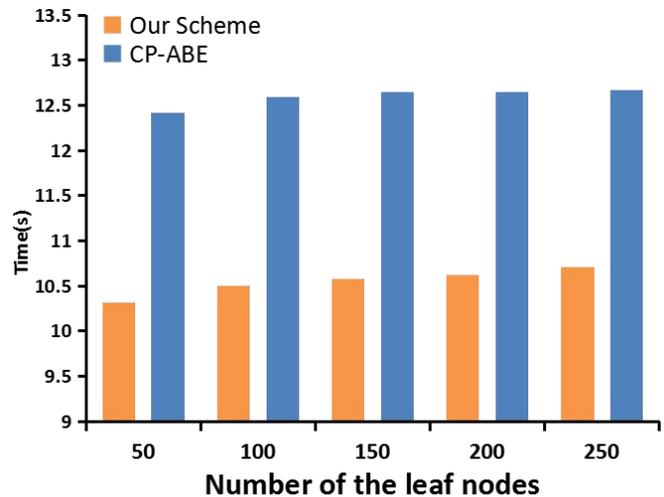

Fig. 4 (c) Comparison of the encryption and transmission time between CP-ABE and our scheme, when the total number of leaves nodes grows.

Finally, with a fixed number of the leaf nodes, the results in Fig. 4 (d) indicate that the total time of our scheme drops slightly as the number of blocks grows.

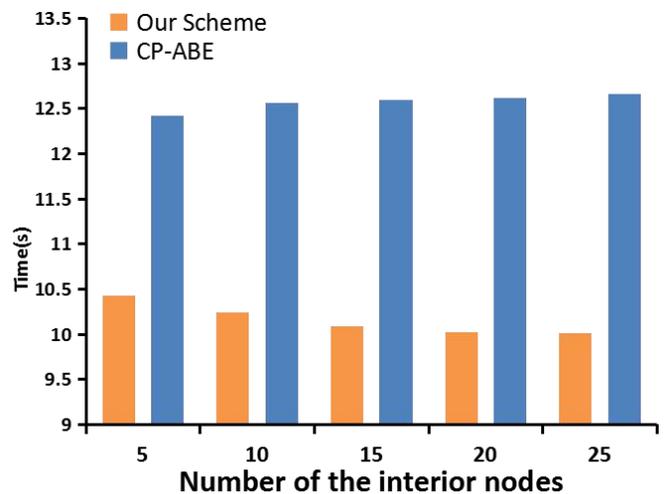

Fig. 4 (d) Comparison of the encryption and transmission time between CP-ABE and our scheme, when the number of blocks grows.

## VI. CONCLUSION

Cloud-IoT techniques are widely deployed in Smart Grid. Huge amount of data is gathered by IoT front-end devices and stored in the back-end cloud servers. However, achieving data security and system efficiency in the data acquisition and transmission process are of great significance and challenging. Existing related schemes cannot deal with this challenging issue well. To tackle with this problem, we propose a secure and efficient data acquisition scheme for Cloud-IoT in smart grid. In the proposed scheme, the large data is partitioned into several blocks, and the blocks are encrypted/decrypted and transmitted in sequence. In addition, we adopt the dual secret sharing scheme, which realizes the privacy-preserving, the data integrity check and the attributes check simultaneously. The analysis shows that the proposed scheme can meet the security requirements of data acquisition in smart grid, and it also reduces response time overhead significantly compared to other popular schemes. The data of the proposed scheme is not uploaded in real time, it is offline before encryption. The research on data timeliness will be our future work.

ACKNOWLEDGMENT

This work is partially supported by Natural Science Foundation of China under grant 61402171, the Fundamental Research Funds for the Central Universities under grant 2016MS29, as well as the US Air Force Research Lab under grant AF16-AT10 and the Qatar National Research Fund under grant NPRP 8-408-2-172.